\documentclass[sn-nature]{sn-jnl}
\usepackage{anyfontsize}
\usepackage[T1]{fontenc}
\usepackage[utf8]{inputenc}
\usepackage{nicefrac}
\usepackage{graphicx}
\usepackage{hyperref}
\usepackage{support-caption}
\usepackage{subcaption}
\usepackage{multirow}%
\usepackage{amsmath,amssymb,amsfonts}%
\usepackage{amsthm}%
\usepackage{mathrsfs}%
\usepackage[title]{appendix}%
\usepackage{xcolor}%
\usepackage{textcomp}%
\usepackage{manyfoot}%
\usepackage{booktabs}%
\usepackage{bookmark}
\usepackage{algorithm}%
\usepackage{algorithmicx}%
\usepackage{algpseudocode}%
\usepackage{listings}
\usepackage{bm}
\usepackage{tikz}
\usepackage{xcolor}
\usetikzlibrary{arrows,decorations.markings}
\usepackage{csquotes}
\usepackage[skins,theorems,most]{tcolorbox}
\tcbset{highlight math style={enhanced, colframe=black!60!black,colback=white,arc=4pt,boxrule=1pt}} 
\usepackage{soul}
\usepackage{empheq}
\usepackage[normalem]{ulem}
\usepackage{geometry}
\usepackage{natbib}

\definecolor{lime}{HTML}{A6CE39}
\DeclareRobustCommand{\orcidicon}{%
	\begin{tikzpicture}
	\draw[lime, fill=lime] (0,0) 
	circle [radius=0.16] 
	node[white] {{\fontfamily{qag}\selectfont \tiny ID}};
	\draw[white, fill=white] (-0.0625,0.095) 
	circle [radius=0.007];
	\end{tikzpicture}
	\hspace{-2mm}
}
\foreach \x in {A, ..., Z}{%
	\expandafter\xdef\csname orcid\x\endcsname{\noexpand\href{https://orcid.org/\csname orcidauthor\x\endcsname}{\noexpand\orcidicon}}
}
\begin{document}

\title[Article Title]{Charged Bardeen
black hole with a cosmological constant and  surrounded by quintessence and a cloud of strings}

\newcommand{\orcidauthorA}{0009-0009-7943-5368}
\newcommand{\orcidauthorB}{0000-0001-7893-0265}
\newcommand{\orcidauthorC}{0000-0001-9284-0549}
\newcommand{\orcidauthorD}{0000-0002-2226-3579}
\newcommand{\orcidauthorE}{0000-0003-4515-9245}

\author*[1]{\fnm{F. F.} \sur{Nascimento}\,\orcidA{}}\email{fran.nice.fisica@gmail.com}

\author[1]{\fnm{V. B.} \sur{Bezerra}\,\orcidB{}}\email{valdir@fisica.ufpb.br}

\author[1]{\fnm{J. M.} \sur{Toledo}\,\orcidC{}}\email{jefferson.m.toledo@gmail.com}

\author[1]{\fnm{P. H.} \sur{Morais}\,\orcidD{}}\email{phm@academico.ufpb.br}

\author[2]{\fnm{J. C.} \sur{Rocha}\,\orcidE{}}\email{julio.rocha@servidor.uepb.edu.br}

\affil*[1]{\orgdiv{Physics Department}, \orgname{Federal University of Para\'iba}, \orgaddress{\city{ Jo\~ao Pessoa}, \postcode{58059-900}, \state{PB}, \country{Brazil}}}

\affil[2]{\orgdiv{Physics Department}, \orgname{State University of Para\'iba}, \orgaddress{\city{Campina Grande}, \postcode{58429-500}, \state{PB}, \country{Brazil}}}


\abstract{Exact solutions which generalize the
Bardeen black hole solution, in the sense that several sources are taken into account, are derived. The most general case corresponds to a charged Bardeen black hole with a cosmological constant and surrounded by quintessence and a cloud of strings. A discussion is presented concerning the Kretschmann scalar and its dependence with the parameters associated to the different sources. Finally, the geodesics and effective potential are analyzed.}

\keywords{}
\maketitle

\section{Introduction}
\label{sec1}

The first steps towards constructing solutions to Einstein's equations without singularities were taken in the sixties of the last century. Sakharov \cite{sakharov1966initial}, for example, from a study of structure formation in a young expanding universe, showed that if the energy density of matter, $\rho$, and the pressure of matter, $p$, are related as, $p=-\rho$, which is the equation of state of the Sitter metric, the agglomeration of baryonic matter does not produce the divergence of $\rho$. That is, with the agglomeration of matter as a result of gravitation, the energy density does not diverge within that formation. But this only occurs in the case where the space-time, in its interior, is of the de Sitter type.

The interesting aspect of de Sitter-type space-time that differs from Minkowski space-time is the addition of a term known as the cosmological constant $\Lambda$. This is the famous constant that Einstein added to his gravitational field equations to obtain a static universe at large scales.

With Hubble and his team's detection of cosmic expansion in the $1920$s \cite{hubble1929relation}, which revealed that the cosmos was no longer regarded as static on huge scales, Einstein was forced to abandon his constant. But, with the measurement of the accelerated expansion of the universe in $1998$ \cite{riess1998observational,riess1999bvri,perlmutter1999measurements}, this constant returns to physics, but now with another role.

The year 1998 was a very important milestone for cosmology. Observations of type Ia supernovae revealed that the universe is expanding at an accelerated rate. Understanding this cosmic acceleration has been one of the most important questions in modern cosmology.

In the context of Einstein's general relativity, accelerated expansion can be explained by introducing into the universe a form of exotic energy with negative pressure, called dark energy, capable of overcoming gravity. Today, the current cosmological model predicts that the universe is composed approximately of $5\%$ of visible matter, $25\%$ of dark matter, and $70\%$ of dark energy \cite{ade2014planck}. Thus, in some cosmological models, the cosmological constant is the cause of the accelerated expansion of the universe, that is, it is the origin of the so-called dark energy.

The de Sitter solution is essential in many areas of physics research, including the study of black holes. Another context is in cosmology when we have the so-called inflationary phase, where the universe had an accelerated expansion right after the big bang, which is described as a period where space-time is almost de Sitter. Anti-de Sitter space, on the other hand, is vital for the, now widely studied, Conformal Field Theory anti-de Sitter Space-time \cite{nastase2007introduction}.

From Sakharov's result, Bardeen \cite{bardeen1968non} constructed the first black hole metric without a singularity, which is spherically symmetric and differs from Schwarzschild's solution \cite{schwarzschild1916uber} in having a mass  $M(r)$, which depends on the radial coordinate. The Bardeen black hole was the first example of a black hole without a singularity. 
Decades after its publication, Bardeen's metric is still being investigated. In \cite{ayon2000bardeen} it is shown that the Bardeen black hole has an origin, that is, one can interpret it as an exact solution of the gravitational field equations.

 Today, there are other regular solutions available in the literature, among them, we cite the Hayward black hole solution \cite{hayward2006formation}, for example, which was constructed  similarly to Bardeen's, with an appropriate function $M(r)$ to describe the formation and evaporation of regular black holes.

In recent years, the research on regular black holes has gone further, and different solutions corresponding to regular black holes were obtained due to the importance of these solutions in understanding these structures \cite{dymnikova1992vacuum,dymnikova2003spherically, ayon1998regular, ayon1999new, sajadi2017nonlinear, balart2014regular, frolov2016notes, bronnikov2001regular, abbas2014geodesic, babichev2020regular, bronnikov2007regular, zhou2012geodesic, wei2015null, molina2021thermodynamics, hayward2006formation, rodrigues2016regular, ghosh2015nonsingular, Ling_2023}. Thus, we think that it is also important
to extend these regular solutions in which a fluid, for example, is encompassing a regular black hole, as well as, others more general and diverse sources \cite{rodrigues2022bardeenclounds, rodrigues2018bardeen, santos2024regular, nascimento2024somebardeen, nascimento2024somehaywardcloundofstrings, lemos2011regular,nascimento2024black}

The paper is organized as follows: In Sec. \ref{sec2} we review the Bardeen black hole solution. In Sec. \ref{sec3}, we   find the solution corresponding to 
a charged Bardeen black hole with a cosmological
constant and surrounded by quintessence and a
cloud of strings.
In Section \ref{sec4}, we calculate the Kretschmann scalar and discuss the results for the general solution as well as for some particular cases.
In Sec. \ref{sec6}, we analyse the geodesics and the effective potential. In Sec. \ref{clonclusion}, we present the concluding remarks. 
In this work, we will use the spherical coordinate system $(t, r, \theta, \phi)$ and  adopt  Planck units, in which the speed of light in vacuum $c$, the gravitational constant $G$, the Boltzmann constant $k_B$, and the reduced Planck constant $\hbar$, are equal to 1. We will also use a metric signature given by $(+---)$.
%
%
\section{Bardeen Black Hole}
\label{sec2}

From Sakharov's result, Bardeen \cite{bardeen1968non} constructed the first black hole metric without a singularity.
The Bardeen's solution has spherical symmetry and differs from Schwarzschild's solution \cite{schwarzschild1916uber} in having a mass that is not a constant $m$, but a function $M(r)$, which depends on the radial coordinate. While in the Schwarzschild solution, the mass is punctual and located in the center of the black hole, in the Bardeen
metric it is spread throughout all space-time. The function $M(r)$ must necessarily make the Bardeen metric, at its core, a de Sitter-type space-time. With this requirement, the space-time inside the black hole is regularized and is then called a regular black hole. 

A spherically symmetric metric describing a black hole can be written as

\begin{equation}
ds^2=f(r) dt^2-\frac{1}{f(r)}dr^2-r^2 d\theta^2-r^2\sin^2\theta d\phi^2,
\label{eq:1.03}
\end{equation}
\noindent
In the case of the Schwarzschild solution in which case the mass of the black hole is constant, $f(r)$ is given by

\begin{equation}
f(r)=1-\frac{2m}{r}.
\label{eq:1.04}
\end{equation}

\noindent A similar algebraic expression can be used to describes Bardeen's regular solution, but now, with a variable mass,

\begin{equation}
f(r)=1-\frac{2M(r)}{r},
\label{eq:1.05}
\end{equation}
with
\begin{equation}
M(r)=\frac{mr^3}{(r^2+q^2)^{3/2}},
\label{eq:1.06}
\end{equation}
where $m$ is interpreted as a mass parameter, while $q$ is taken as a type of charge \cite{ayon2000bardeen}. We can observe in Eq. (\ref{eq:1.06})  that if $r\rightarrow \infty$, $M(r)\rightarrow m$. Thus, very far from the black hole, the Bardeen solution is similar to the Schwarzschild one. For $r \rightarrow 0$, we can write $M(r) \approx \frac{mr^3}{q^3}$, and then $f(r)\approx 1-Cr^2$, with $C=\frac{2m}{q^3}$ being a positive constant. Therefore,  in this limit, the space-time metric with $f(r)$ given by Eq. (\ref{eq:1.05}) and (\ref{eq:1.06}) is similar to de Sitter space-time. Thus, the Bardeen black hole has an internal core with behavior similar to de Sitter metric \cite{bardeen1968non}.

 Thus, this solution has no singularity at its center, and for this reason, it is know as regular black hole, due to the fact that the spacetime has no singularity in any point, including, obviously, 
 $r=0$, differently from the Schwarzschild metric which diverges at the origin of the coordinate system, independently from  the system of coordinates used. 
 
There are general approaches to test the existence of singularities. 
One of them uses the behavior of some invariant constructed with the curvature tensor, $R_{\mu\nu}$ or $R_{\mu\nu\sigma\rho}$, to  test the singular nature of a spacetime. The resulting quantity permits to perform a local and simple investigation about the existence of singularities of the manifold considered.
Among these quantities, we  will consider
 the Kretschmann scalar, $K$ which is given by $K = R_{\mu\nu\sigma\rho}R^{\mu\nu\sigma\rho}$.

The Kretschmann scalar for the Bardeen metric is given by 

\begin{equation}
\begin{aligned}
K=\frac{12m^2}{(r^2+q^2)^7}(8q^8-4q^6r^2+47q^4r^4-12q^2r^6+4r^8),
\label{eq:1.011}
\end{aligned}
\end{equation}

\noindent whose limits are the following:

\begin{equation}
\lim_{r\rightarrow 0} K=96\frac{m^2}{q^6},
\label{eq:1.012}
\end{equation}

\noindent and

\begin{equation}
\lim_{r\rightarrow \infty} K=0.
\label{eq:1.0012}
\end{equation}

It is worth calling attention to the fact that for Schwarzschild black hole, the Kretschmann scalar is singular at $r=0$, while in the case of Bardeen black hole  , the Kretschmann scalar is finite as $r\rightarrow 0$. Consequently, the Bardeen black hole does not exhibit a curvature singularity at the origin, according to the approach based on the analysis of the Kretschmann scalar.

\subsection{Nonlinear electrodynamics scenario}

The source of Bardeen's black hole solution is a monopole charge, $q$, of a self-gravitating magnetic field in nonlinear electrodynamics \cite{ayon2000bardeen}. This means that the magnetic monopole arises from a nonlinear electrodynamic field coupled to general relativity, whose action can be expressed as

\begin{equation}
S=\int \left(\frac{1}{16\pi}R-\frac{1}{4\pi}\mathcal{L}(F)\right)dv,
\label{eq:1.015}
\end{equation}

\noindent where $R$ is the scalar curvature and $\mathcal{L}(F)$
is the Lagrangian density
corresponding to the source of the
Bardeen black hole
in the framework of the nonlinear electrodynamics \cite{ayon2000bardeen}, and is written as

\begin{equation}
\mathcal{L}(F)=\frac{3}{2sq^2}\left(\frac{\sqrt{2q^2F}}{1+\sqrt{2q^2F}}\right)^{5/2},
\label{eq:1.016}
\end{equation}

where $s$ is given by

\begin{equation}
s\equiv \frac{|q|}{2m}.
\label{eq:1.017}
\end{equation}

Thus, the Bardeen black hole solution is obtained by solving the Einstein equations for an energy-momentum tensor associated to a nonlinear electrodynamic field.
It is worth noting that when the second term of Eq. (\ref{eq:1.015}) is zero, the Einstein equations are obtained in a vacuum and correspond to the Schwarzschild black hole.

\section{Charged Bardeen black hole with a cosmological
constant and surrounded by quintessence and a
cloud of strings.}
\label{sec3}

In this section we obtain
the solution  corresponding to a static, charged Bardeen
black hole with a cosmological constant and surrounded by
quintessence and a cloud of strings. To do this,
Einstein's equations coupled with a nonlinear electromagnetics field are solved taking into account the mentioned sources.
 
%
%

Taking into account the 
presence of the cosmological constant,
the Einstein-Hilbert
action turns into

\begin{eqnarray}\label{actioEH}
S_{EH}=\frac{1}{2\kappa^{2}}\int{d^{4}x\,\sqrt{-g}\left(R+2\Lambda\right)}\,\mbox{.}
\end{eqnarray}

Concerning the contribution to the action
arising from
others sources and from the coupling with the nonlinear electromagnetic field,
which we are calling $S_{CNLS}$,
where $CNL$ means the coupling 
between the nonlinear electromagnetic field and the quintessence, which appears in the first term of $S_{CNLS}$, showed bellow, and $S$ refers to
the contribution associated with the electromagnet field and the cloud of strings,
given by the second and third terms, respectively. Thus, the action
$S_{CNLS}$  can be written as

\begin{eqnarray}\label{actioM}
\begin{aligned}
S_{CNLS}=&\int{d^{4}x\,\sqrt{-g}\,L(F)}\, + \int{d^{4}x\,\sqrt{-g}\,L_{E}}+\int{d^{4}x\,\sqrt{-g}\,L_{CS}}.
\end{aligned}
\end{eqnarray}

\noindent The Lagrangian corresponding to the nonlinear electromagnetic field coupled to quintessence, $L(F)$, is given by \cite{rodrigues2022bardeen,ayon2000bardeen,kiselev2003quintessence}

\begin{eqnarray}\label{L(F)}
L(F)=\frac{24\sqrt{2}mq^2}{\kappa^2\left(\sqrt{\frac{2q^2}{F}}+2q^2\right)^{5/2}}-\frac{3\omega_q \alpha\left(\frac{2F}{q^2}\right)^{\frac{3(\omega_q+1)}{4}}}{\kappa^2},
\end{eqnarray}

\noindent with $F$ defined by $F=F^{\mu\nu}F_{\mu\nu}/4$. The only non-zero component of the Maxwell-Faraday tensor, $F_{\mu\nu}$, for a spherically symmetric and  magnetically charged spacetime is \cite{ayon2000bardeen,bronnikov2001regular,zhang2018first,plebanski2006introduction}:

\begin{eqnarray}\label{F23}
F_{23}=q \sin{\theta},
\end{eqnarray}
\noindent where $q$ is the magnetic charge and the scalar $F$ is given by
\begin{eqnarray}\label{F}
F=\frac{q^2}{2r^4}.
\end{eqnarray}

\noindent The Lagrangian of the electromagnetic field $L_E$ is  \cite{d2022introducing}:

\begin{eqnarray}\label{L_E}
L_{E}=-\frac{1}{4}F_{\alpha\beta}F^{\alpha\beta},
\end{eqnarray}

\noindent with

\begin{eqnarray}\label{F01}
F_{01}=F_{10}=-F^{01}=E_{r}=\frac{Q}{r^2}.
\end{eqnarray}

\noindent For stringlike objects, the action of Nambu-Goto is given by \cite{letelier1979clouds,nascimento2024somehaywardcloundofstrings}

\begin{eqnarray}\label{S_{CS}}
S_{CS}=\int{d^{4}x\,\sqrt{-g}\,L_{CS}}\,=\int (-\gamma)^{1/2}\mathcal{M}{d\lambda^{0}d\lambda^{1}},
\end{eqnarray}

\noindent whose Lagrangian $L_{CS}$ for the cloud of strings is given by \cite{letelier1979clouds}:

\begin{eqnarray}\label{L_{CS}}
L_{CS} =\mathcal{M}\sqrt{-\gamma}= \mathcal{M}\left(-\frac{1}{2}\Sigma^{\mu\nu}\Sigma_{\mu\nu}\right)^{1/2}.
\end{eqnarray}

For the general case, namely, charged Bardeen black hole with a cosmological constant and surrounded by  quintessence and a cloud of strings,  the total action is written as

\begin{eqnarray}\label{total action}
\begin{aligned}
S=&S_{EH}+S_{CNLS}\\=&\frac{1}{2\kappa^{2}}\int{d^{4}x\sqrt{-g}\left(R+2\Lambda\right)}+\int{d^{4}x\sqrt{-g}L(F)}\\
+&\int{d^{4}x\sqrt{-g}L_{E}}+\int{d^{4}x\sqrt{-g}L_{CS}}.
\end{aligned}
\end{eqnarray}

Thus, by varying the action (\ref{total action}) with respect to the metric, we obtain

\begin{eqnarray}\label{equationfield}
R_{\mu\nu}-\frac{1}{2}g_{\mu\nu}R -\Lambda g_{\mu\nu} = \kappa^{2}(T_{\mu\nu}^{CNL} + T_{\mu\nu}^{E} + T_{\mu\nu}^{CS}\,)\mbox{,}
\end{eqnarray}

\noindent with the energy-momentum tensors
given by \cite{rodrigues2022bardeen,d2022introducing,letelier1979clouds}:

\begin{eqnarray}\label{2.19}
T_{\mu\nu}^{CNL} = g_{\mu\nu}L(F) -\frac{\partial L}{\partial F}F_{\mu}^{\;\alpha}F_{\nu\alpha},
\end{eqnarray}

\begin{eqnarray}\label{2.21}
T_{\mu\nu}^{E} = \frac{1}{\kappa^2}\left(-2g^{\alpha \beta}F_{\mu \alpha}F_{\nu \beta}+\frac{1}{2}g_{\mu \nu}F_{\alpha \beta}F^{\alpha \beta}\right),
\end{eqnarray}

\begin{eqnarray}\label{2.22}
T_{\mu\nu}^{CS} = \frac{\rho\Sigma_{\mu}^{\;\beta}\Sigma_{\beta\nu}}{\kappa^2(-\gamma)^{1/2}}.
\end{eqnarray}.

The nonzero
components of these tensors, namely,
the components $(0,0)$ and $(2,2)$, are given by

\begin{equation}\label{T00BK}
\begin{aligned}
T_{00}^{CNL} =&\frac{f(r)}{\kappa^2}\left(\frac{6mq^2}{\left(r^2+q^2\right)^{5/2}}-3\omega_q \alpha r^{r^{-3\omega_q -3}}\right).
\end{aligned}
\end{equation}

\begin{equation}\label{T22BK}
\begin{aligned}
T_{22}^{CNL} =&\frac{1}{\kappa^2}\left(-\frac{3 m q^2 r^2 \left(2 q^2-3 r^2\right)}{\left(q^2+r^2\right)^{7/2}}-\frac{3 \alpha \omega_q  (3 \omega_q +1) r^{r^{-3 \omega_q -1}}}{2}\right).
\end{aligned}
\end{equation}

\begin{eqnarray}\label{T00E}
\begin{aligned}
T_{00}^{E} = \frac{f(r)}{\kappa^2}\frac{Q^2}{r^4},\;\;\;\;\;\;T_{22}^{E} =\frac{1}{\kappa^2}\frac{Q^2}{r^2}.
\end{aligned}
\end{eqnarray}


\begin{eqnarray}\label{T00CS}
\begin{aligned}
T_{00}^{CS} =\frac{f(r)}{\kappa^2}\frac{a}{r^2},\;\;\;\;\;\;T_{22}^{CS} =0.
\end{aligned}
\end{eqnarray}

\noindent 
Taking these results
for the energy-momentum tensors
into account, 
 Eq. (\ref{equationfield})
 reduces to the following differential equations

\begin{equation}
\begin{aligned}  
&-\frac{a}{r^2}-\frac{f'(r)}{r}-\frac{f(r)}{r^2}+\frac{1}{r^2}-\Lambda -\frac{6 m q^2}{\left(q^2+r^2\right)^{5/2}}-\frac{Q^2}{r^4}+3\alpha  \omega_q  r^{-3 (\omega_q +1)}=0,
\label{edo1}
\end{aligned}
\end{equation}

\begin{equation}
\begin{aligned}  
&\frac{1}{2} r^2 f''(r)+r f'(r)+\frac{3 m q^2 r^2 \left(2 q^2-3 r^2\right)}{\left(q^2+r^2\right)^{7/2}}\\
&-\frac{Q^2}{r^2}+\Lambda  r^2+\frac{3}{2} \alpha  \omega_q  (3 \omega_q +1) r^{-3 \omega_q -1}=0.
\label{edo2}
\end{aligned}
\end{equation}

\noindent Multiplying Eq. (\ref{edo1}) by $r^2$ and Eq. (\ref{edo2}) by $2$, and adding the results, we get: 

\begin{equation}
\begin{aligned}
&1-a-f(r)+r f'(r)+r^2 f''(r)\\
&+\frac{6 m q^2 r^2 \left(2 q^2-3 r^2\right)}{\left(q^2+r^2\right)^{7/2}}-\frac{6 m q^2 r^2}{\left(q^2+r^2\right)^{5/2}}\\
&-\frac{3 Q^2}{r^2}+3 \alpha  \omega_q  (3 \omega_q +2) r^{-3 \omega_q -1}+\Lambda  r^2=0.
\end{aligned}
\label{edo3}
\end{equation}

\noindent Solving
Eq. (\ref{edo3}),
we obtain the following result

\begin{equation}
\begin{aligned}
f(r)=&1-a-\frac{2 m r^2}{\left(q^2+r^2\right)^{3/2}}- \alpha  r^{-3 \omega_q -1}+\frac{Q^2}{r^2}-\frac{\Lambda  r^2}{3}+\frac{C_1}{r}+C_2 r,
\label{solution}
\end{aligned}
\end{equation}

\noindent where $C_{1}$ and $C_{2}$ are arbitrary constants whose values are fixed by imposing some conditions which should be obeyed by the physical system. 
In one  of these conditions 
 we assume that $C_{1}=0$, in order to recover Bardeen’s solution, when  all additional sources
are absent.
It is worth calling
attention to the fact that the studies we are performing consider as one of the sources the quintessence, and thus the parameter $\omega_q$ assumes the value $-2/3$. Therefore, the fourth and the eighth terms in Eq. (\ref{solution}) can be combined, given a new effective arbitrary constant. 
Then, taking these results into acount,
the function $f (r)$ can be written in a simplified form without the last two terms. Thus, using this simplified form,  the metric corresponding to a charged Bardeen black hole
with a cosmological constant and surrounded by quintessence and a cloud of strings, in a non-linear electrodynamics scenario coupled to Einstein gravity, is given by

\begin{eqnarray}\label{eq:1.31}
\begin{aligned}
&ds^{2}=
\\&+\left(1-a-\frac{2 m r^2}{\left(q^2+r^2\right)^{3/2}}-\frac{\alpha}{r^{3 \omega_q+1}}+\frac{Q^2}{r^2}-\frac{\Lambda r^2 }{3} \right)dt^{2}\\
&-\left(1-a-\frac{2 m r^2}{\left(q^2+r^2\right)^{3/2}}-\frac{\alpha}{r^{3 \omega_q+1}}+\frac{Q^2}{r^2}-\frac{\Lambda r^2}{3} \right)^{-1}dr^{2}\\
&- r^{2}d\theta^{2} - r^{2}\sin^{2}\theta d\phi^{2}.
\end{aligned}
\end{eqnarray}

\noindent The spacetimes that can be recovered from Eq. (\ref{eq:1.31}) are described in the Table \ref{table}. 

\begin{table}
\caption{Space-time that can be recovered from the Eq. (\ref{eq:1.31})}
\begin{tabular}{lcclccclcccclcl}
\hline
&         BH PARAMETERS                  &     SPACE-TIME           &\\\hline
&          $a=\alpha=Q=\Lambda=0$        &     Bardeen              &\\
&         $a=\alpha=\Lambda=0$           &     Bardeen-Reissner-Nordström&\\
&          $\alpha=Q=\Lambda=0$          &     Bardeen-Letelier&\\
&         $a=Q=\Lambda=0$                &     Bardeen-Kiselev&\\
&         $a=\alpha=Q=0$                 &     Bardeen-AdS&\\
&         $Q=\Lambda=0$                  &     Bardeen-Kiselev-Letelier&\\
&          $Q=0$                         &     Bardeen-AdS-Kiselev-Letelier&\\\hline
\end{tabular}
\label{table}
\end{table}

%
%
\section{Kretschmann Scalar analysis}
\label{sec4}

If we want to verify the existence of singularities in this space-time, one possible route is to calculate the Kretschmann scalar. Thus, for the metrics given by Eq. (\ref{eq:1.31}), the Kretschmann scalar is given by

\begin{equation}
\begin{aligned}
K&=R_{\alpha\beta\mu\nu}R^{\alpha\beta\mu\nu}
\\
&=\frac{4 a^2}{r^4}+\frac{16 a m}{r^2 \left(q^2+r^2\right)^{3/2}}+8 a\alpha r^{-3\omega_q-5} +\frac{8 \Lambda ^2}{3}+\frac{8 a \Lambda }{3 r^2}
\\
&+\frac{12 m^2 \left(8 q^8-4 q^6 r^2+47 q^4 r^4-12 q^2 r^6+4 r^8\right)}{\left(q^2+r^2\right)^7}+\frac{56 Q^4}{r^8}
\\
&+4\alpha \Lambda\omega_q (3\omega_q-1) r^{-3 (\omega_q+1)}-12 \alpha Q^2 \left(9 \omega_q^2+13 \omega_q+4\right) r^{-3\omega_q-7}
\\
&+\frac{24 m Q^2 \left(11 q^2-4 r^2\right)}{r^2 \left(q^2+r^2\right)^{7/2}}+\frac{12\alpha m r^{-3 (\omega_q+1)} h(r)}{\left(q^2+r^2\right)^{7/2}}-\frac{8 a Q^2}{r^6}
\\
&+3\alpha^2 \left(27 \omega_q^4+54 \omega_q^3+51\omega_q^2+20\omega_q+4\right) r^{-6 (\omega_q+1)}
,
\label{eq:1.32}
\end{aligned}
\end{equation}
where
\begin{eqnarray}
    \begin{aligned}
    h(r) &= 2 r^4 \left(3 \omega_q^2+5 \omega_q+2\right) -q^2 r^2 \left(33 \omega_q^2+37 \omega_q+6\right) + 2 q^4 (3\omega_q^{2}-\omega_q) \nonumber \\
    &+ 8 \Lambda  m \left(4 q^4-q^{2}r^2\right).
    \end{aligned}
\end{eqnarray}

\noindent Now, let us investigate  the limit of the Kretschmann scalar when $r\rightarrow 0$ and $r\rightarrow \infty$.

\begin{figure}[!ht]
  \centering
  \begin{subfigure}[b]{0.47\textwidth}
    \includegraphics[width=\textwidth]{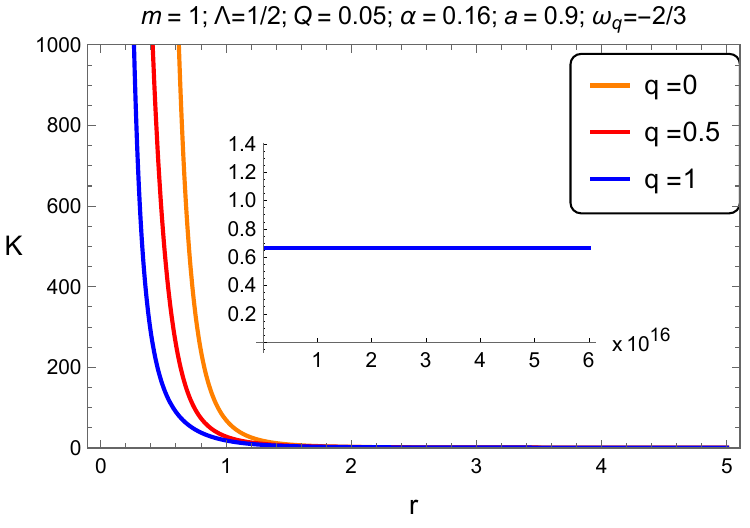}
    \caption{}\label{im1a}
  \end{subfigure}
  \vspace{.5cm}
  \begin{subfigure}[b]{0.47\textwidth}
    \includegraphics[width=\textwidth]{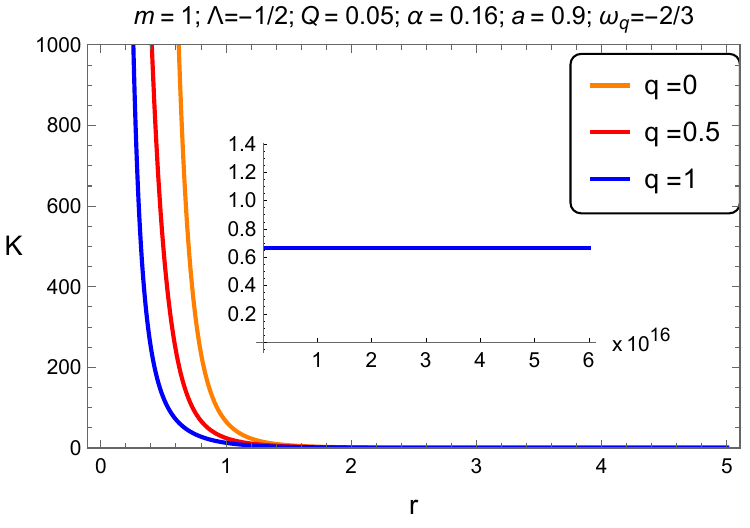}
    \caption{}\label{im1b}
  \end{subfigure}
   \begin{subfigure}[b]{0.47\textwidth}
    \includegraphics[width=\textwidth]{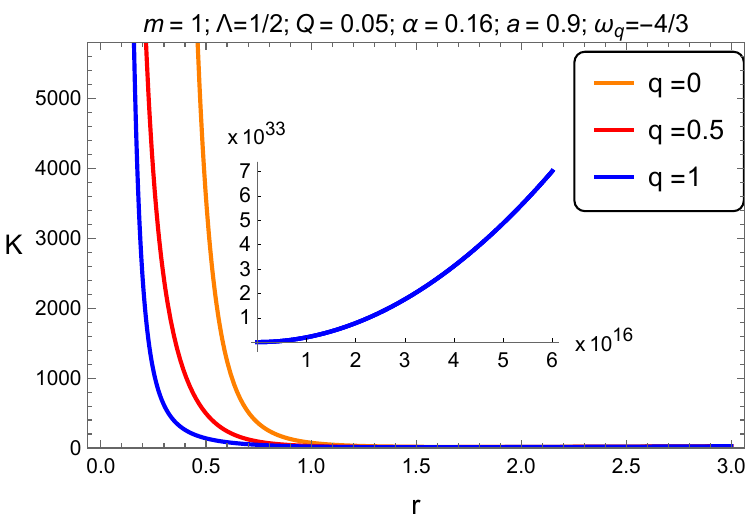}
    \caption{}\label{im1c}
  \end{subfigure}
   \begin{subfigure}[b]{0.47\textwidth}
    \includegraphics[width=\textwidth]{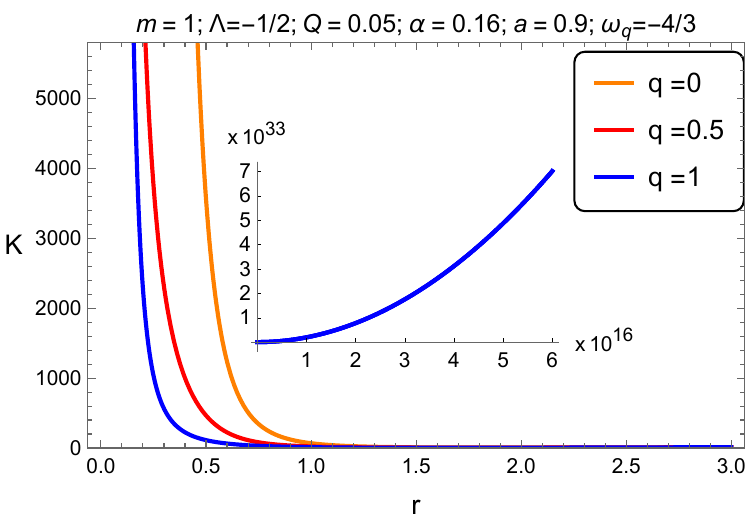}
    \caption{}\label{im1d}
  \end{subfigure}
 \caption{Kretschmann scalar referring to the 
 charged Bardeen  black hole with a cosmological constant and surrounded by
 quintessence and a cloud of strings, for different values of the parameters $q$, $\Lambda$, and $\omega_q$.}
  \label{im1}
\end{figure}

\noindent In the quintessence energy regime  $-1<\omega_q<-1/3$, we have the following limits:
 
\begin{equation}
\lim_{r\rightarrow 0}K=\infty\,\,\,\mbox{e}\,\,\,\lim_{r\rightarrow \infty}K=\frac{8 \Lambda ^2}{3}.
\label{eq:1.39}
\end{equation}

\noindent In addition, let us consider the energy regime corresponding to what is termed
phantom, in which case $\omega_q=-4/3$, thus,  we get the following limits: 

\begin{equation}
\lim_{r\rightarrow 0}K=\infty\,\,\,\mbox{e}\,\,\,\lim_{r\rightarrow \infty}K=\infty.
\label{eq:1.35}
\end{equation}

Concluding, we can say that the curvature is finite, when $r\rightarrow \infty$ and $-1<\omega_q<-1/3$,  and singular at the origin.  Otherwise, for $\omega_q=-4/3$, the solution is singular, in both limits, as can be seen in Fig. \ref{im1}.

\subsection{Bardeen black hole}

The Bardeen
black hole is obtained by setting $\alpha=0$, $\Lambda=0$, $Q=0$ and $a=0$ in Eq. (\ref{eq:1.31})

\begin{equation}
\begin{aligned} 
ds^2=&\left(1-\frac{2 m r^2}{\left(q^2+r^2\right)^{3/2}}\right)dt^2
-\left(1-\frac{2 m r^2}{\left(q^2+r^2\right)^{3/2}}\right)^{-1}dr^2-r^2 d\Omega^2,
\label{eq:1.41}
\end{aligned}
\end{equation}

\noindent whose Kretschmann scalar is

\begin{equation}
K=\frac{12 m^2 \left[8 q^8-4 q^6 r^2+47 q^4 r^4-12 q^2 r^6+4 r^8\right]}{\left(q^2+r^2\right)^7},
\label{eq:1.42}
\end{equation}

 \noindent with
 
\begin{equation}
\lim_{r\rightarrow 0}K=\frac{96 m^2}{q^6}\,\,\,\mbox{and}\,\,\,\lim_{r\rightarrow \infty}K=0.
\label{eq:1.43}
\end{equation}

\begin{figure}[h!]
  \centering
    \includegraphics[scale=0.7]{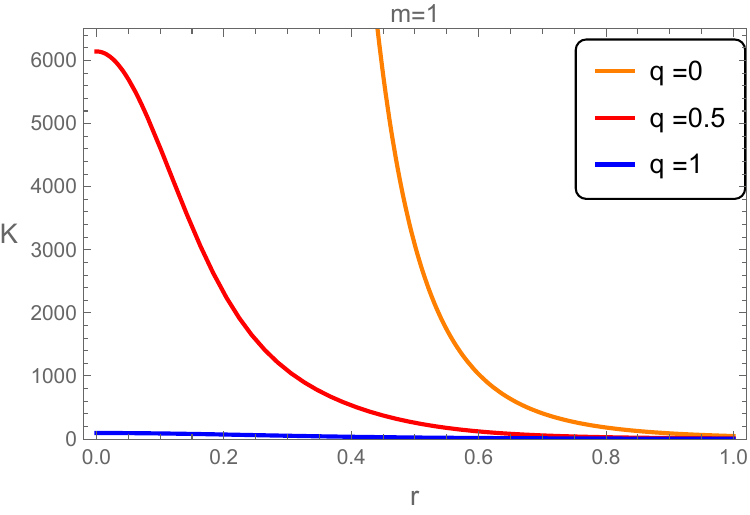}
  \caption{Kretschmann scalar referring to the Bardeen black hole for different values of $q$.}
  \label{im2}
\end{figure}

As expected, the analysis of the Kretschmann scalar shows us that the Bardeen black hole is regular, that is, it has no singularity at the origin $r=0$ and is asymptotically flat. This behavior is described in Fig. \ref{im2}.

\subsection{Bardeen black hole surrounded by quintessence}

By setting $\Lambda=0$, $Q=0$ and $a=0$ in Eq. (\ref{eq:1.31}), we obtain the Bardeen black hole surrounded by quintessence (Bardeen-Kiselev black hole), whose metric reads

\begin{equation}
 \begin{aligned}
ds^2&=\left(1-\frac{2 m r^2}{\left(q^2+r^2\right)^{3/2}}-\frac{\alpha}{r^{3 \omega _q+1}}\right)dt^2
\\
&-\left(1-\frac{2 m r^2}{\left(q^2+r^2\right)^{3/2}}-\frac{\alpha}{r^{3 \omega _q+1}}\right)^{-1}dr^2-r^2 d\Omega^2,
\label{eq:1.45}
\end{aligned} 
\end{equation}

\noindent with Kretschmann scalar given by

\begin{equation}
\begin{aligned}
K&=\frac{12 m^2 \left(8 q^8-4 q^6 r^2+47 q^4 r^4-12 q^2 r^6+4 r^8\right)}{\left(q^2+r^2\right)^7}
\\
&+3 \alpha^2 \left(27 w_q^4+54 w_q^3+51 w_q^2+20 w_q+4\right) r^{-6 (w_q+1)}
\\
&+\frac{12\alpha m r^{-3 (w_q+1)} \left[2 q^4 w_q (3 w_q-1)\right]}{\left(q^2+r^2\right)^{7/2}}
\\
&+\frac{12\alpha m r^{-3 (w_q+1)} \left[-q^2 \left(33 w_q^2+37 w_q+6\right) r^2\right]}{\left(q^2+r^2\right)^{7/2}}
\\
&+\frac{12\alpha m r^{-3 (w_q+1)} \left[+2 \left(3 w_q^2+5 w_q+2\right) r^4\right]}{\left(q^2+r^2\right)^{7/2}}.
\label{eq:1.46}
\end{aligned}
\end{equation}

\noindent In the quintessence energy regime, $\omega_q=-2/3$, we have the following limits:
 
\begin{equation}
\lim_{r\rightarrow 0}K=\infty\,\,\,\mbox{e}\,\,\,\lim_{r\rightarrow \infty}K=0.
\label{eq:1.51}
\end{equation}

\noindent In the phantom
energy regime, namely, for $\omega_q=-4/3$, we get the following limits: 

\begin{equation}
\lim_{r\rightarrow 0}K=\frac{96 m^2}{q^6}\,\,\,\mbox{e}\,\,\,\lim_{r\rightarrow \infty}K=\infty.
\label{eq:1.53}
\end{equation}

\begin{figure}[h!]
  \centering
  \begin{subfigure}[b]{0.47\textwidth}
    \includegraphics[width=\textwidth]{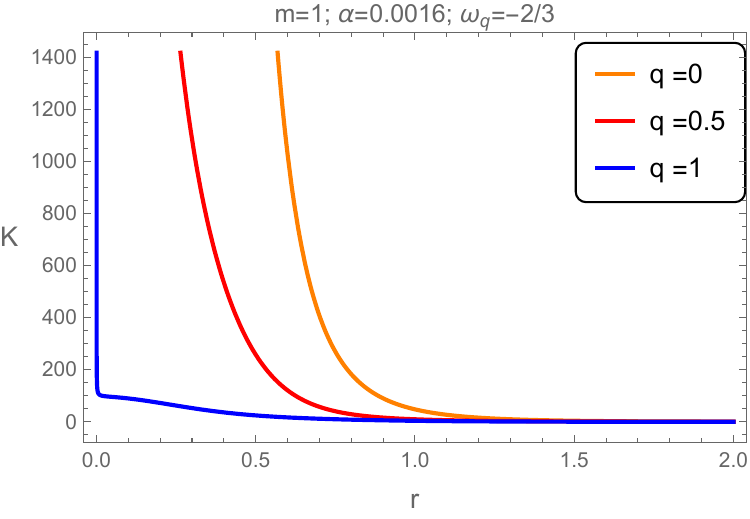}
    \caption{}\label{im3a}
  \end{subfigure}
  \vspace{.5cm}
  \begin{subfigure}[b]{0.47\textwidth}
    \includegraphics[width=\textwidth]{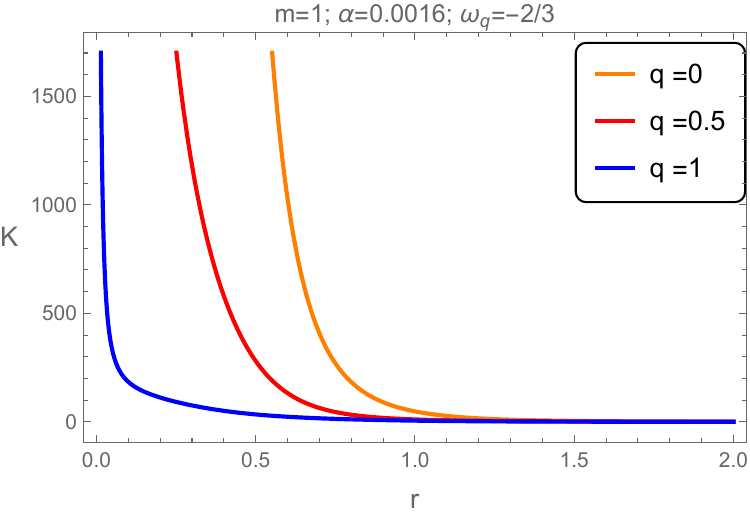}
    \caption{}\label{im3b}
  \end{subfigure}
   \begin{subfigure}[b]{0.47\textwidth}
    \includegraphics[width=\textwidth]{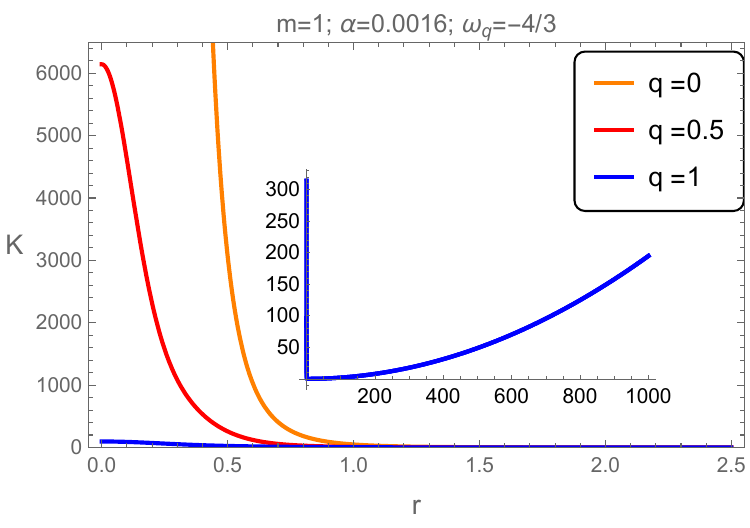}
    \caption{}\label{im3c}
  \end{subfigure}
   \begin{subfigure}[b]{0.47\textwidth}
    \includegraphics[width=\textwidth]{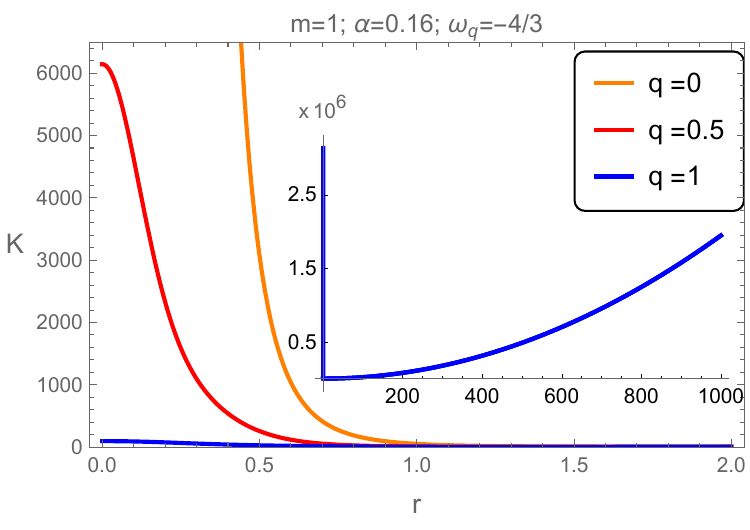}
    \caption{}\label{im3d}
  \end{subfigure}
 \caption{Kretschmann scalar referring to the Bardeen black hole 
 surrounded by quintessence
 for different values of $q$, $\omega_q$, and $\alpha$.}
  \label{im3}
\end{figure}

Therefore, the analysis of the Kretschmann scalar, whose behavior is described in Fig. \ref{im3}, shows that:

\begin{itemize}
\item the Bardeen black hole surrounded by quintessence is no longer regular for $\omega_q=-2/3$ , that is, it has a singularity at the origin $r=0$; however, it is asymptotically flat.
\item the Bardeen black 
hole with a phantom is regular for $\omega_q=-4/3$, that is, it has no singularity at the origin $r=0$; however, it is not asymptotically flat when $r\rightarrow \infty$.
\end{itemize}

\subsection{Bardeen black hole with Cosmological Constant}

The Bardeen black hole with the cosmological constant (Bardeen AdS black hole) is obtained by taking $\alpha=0$, $Q=0$ and $a=0$ in Eq. (\ref{eq:1.31}):

\begin{equation}
 \begin{aligned}
ds^2&=\left(1-\frac{2 m r^2}{(q^2+r^2)^{3/2}}-\frac{1}{3}\Lambda  r^2\right)dt^2\\
&-\left(1-\frac{2 m r^2}{(q^2+r^2)^{3/2}}-\frac{1}{3}\Lambda  r^2\right)^{-1}dr^2-r^2 d\Omega^2,
\label{eq:1.55}
\end{aligned} 
\end{equation}

\noindent whose Kretschmann scalar is

\begin{equation}
\begin{aligned}
K&=\frac{8 \Lambda ^2}{3}+\frac{8 q^2 \Lambda  m \left(4 q^2-r^2\right)}{\left(q^2+r^2\right)^{7/2}}
\\
&+\frac{12 m^2 \left(8 q^8-4 q^6 r^2+47 q^4 r^4-12 q^2 r^6+4 r^8\right)}{\left(q^2+r^2\right)^7}.
\label{eq:1.56}
\end{aligned}
\end{equation}

 \noindent Calculating the limits of the Kretschmann scalar as $r\rightarrow 0$ and $r\rightarrow \infty$, we obtain:
 
\begin{equation}
\lim_{r\rightarrow 0} K=\frac{8 \Lambda ^2}{3}+\frac{32 \Lambda  m}{q^3}+\frac{96 m^2}{q^6},
\label{eq:1.57}
\end{equation}

\begin{equation}
\lim_{r\rightarrow \infty} K=\frac{8\Lambda^2}{3}.
\label{eq:1.58}
\end{equation}

\begin{figure}[h!]
  \centering
  \begin{subfigure}[b]{0.47\textwidth}
    \includegraphics[width=\textwidth]{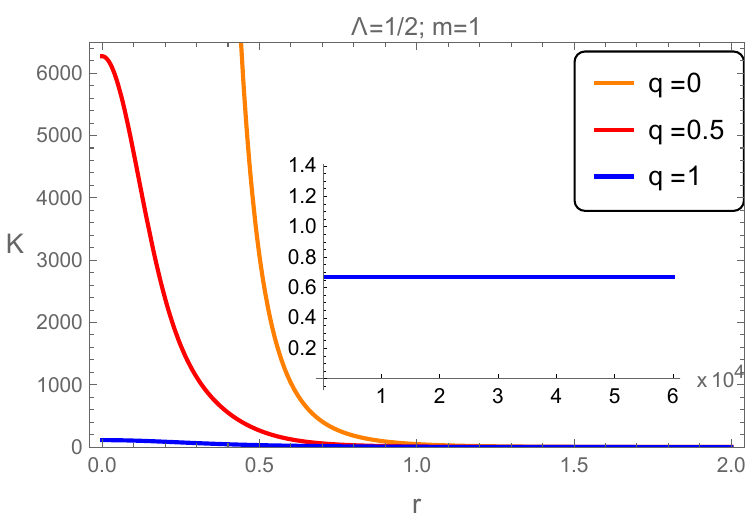}
    \caption{}\label{im4a}
  \end{subfigure}
  \vspace{.5cm}
  \begin{subfigure}[b]{0.47\textwidth}
    \includegraphics[width=\textwidth]{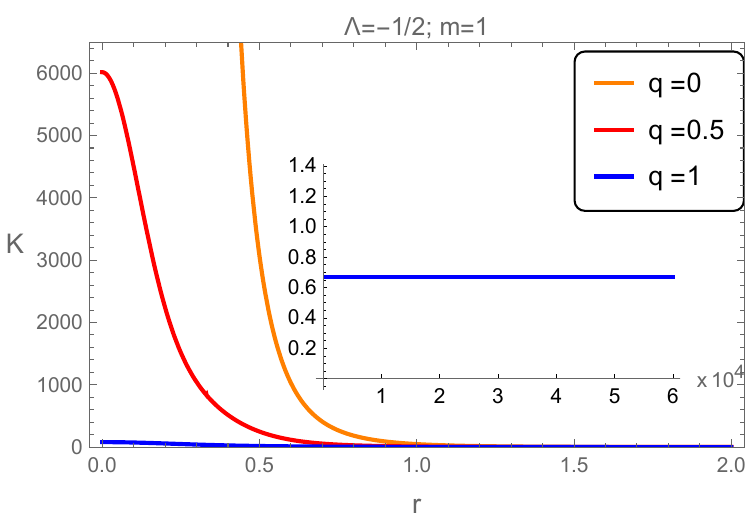}
    \caption{}\label{im4b}
  \end{subfigure}
\caption{Kretschmann scalar referring to the Bardeen black hole with a cosmological constant for different values of $q$ and $\Lambda$.}
  \label{im4}
\end{figure}

Our results indicate that adding a cosmological constant to the Bardeen black hole leaves the regularity of the metric unchanged, a conclusion that is also illustrated in Fig. \ref{im4}.

\subsection{Bardeen black hole with electromagnetic field}

The Bardeen black hole with electromagnetic field (Bardeen-Reissner-Nordström black hole) can be obtained by setting $\alpha=0$, $\Lambda=0$ and $a=0$ in Eq. (\ref{eq:1.31}). The resulting metric is written as

\begin{equation}
 \begin{aligned}
ds^2&=\left(1-\frac{2 m r^2}{(q^2+r^2)^{3/2}}+\frac{Q^2}{r^2}\right)dt^2\\
&-\left(1-\frac{2 m r^2}{(q^2+r^2)^{3/2}}+\frac{Q^2}{r^2}\right)^{-1}dr^2-r^2 d\Omega^2.
\label{eq:1.59}
\end{aligned} 
\end{equation}

\noindent Concerning the Kretschmann scalar, we have the following results

 \begin{equation}
\begin{aligned}
K&=\frac{56 Q^4}{r^8}+\frac{24 m Q^2 \left(11 q^2-4 r^2\right)}{r^2 \left(q^2+r^2\right)^{7/2}}
\\
&+\frac{12 m^2 \left(8 q^8-4 q^6 r^2+47 q^4 r^4-12 q^2 r^6+4 r^8\right)}{\left(q^2+r^2\right)^7},
\label{eq:1.60}
\end{aligned}
\end{equation}

 \noindent whose limits are

\begin{equation}
\lim_{r\rightarrow 0}K=\infty\,\,\,\mbox{and}\,\,\,\lim_{r\rightarrow \infty}K=0.
\label{eq:1.61}
\end{equation}

\begin{figure}[h!]
  \centering
  \begin{subfigure}[b]{0.47\textwidth}
    \includegraphics[width=\textwidth]{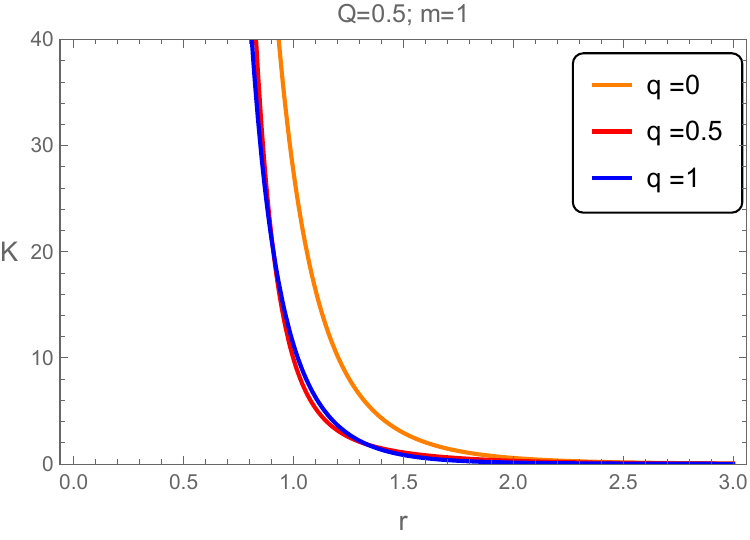}
    \caption{}\label{im5a}
  \end{subfigure}
  \vspace{.5cm}
  \begin{subfigure}[b]{0.47\textwidth}
    \includegraphics[width=\textwidth]{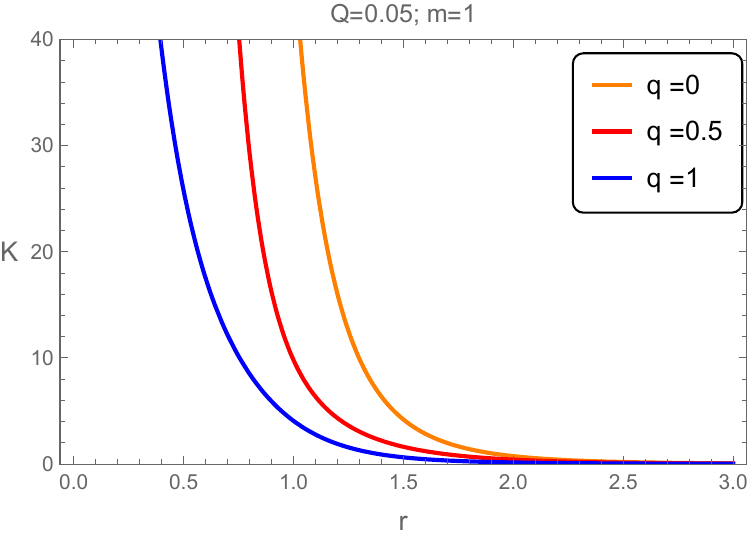}
    \caption{}\label{im5b}
  \end{subfigure}
\caption{Kretschmann scalar referring to the Bardeen black hole with an electromagnetic field for different values of $q$ and $Q$.}
  \label{im5}
\end{figure}

In the Fig. \ref{im5}, we can conclude that the inclusion of the electric charge in Bardeen's metric destroys the regularity of the metric by introducing a singularity at the origin.  

\subsection{Bardeen black hole  with a cloud of strings}

The Bardeen black hole with a cloud of strings (Bardeen-Letelier black hole) can be obtained from Eq. (\ref{eq:1.31}) 
by setting $\alpha=0$, $\Lambda=0$ and $Q=0$.
We also can obtain a known
result
of the literature \cite{rodrigues2022bardeenclounds},
by  setting
$a=0$ and $Q=0$ and $\Lambda$ different from zero.
Thus, the Bardeen black hole solution with a cloud of strings is written as
\begin{equation}
 \begin{aligned}
ds^2&=\left(1-a-\frac{2 m r^2}{(q^2+r^2)^{3/2}}\right)dt^2\\
&-\left(1-a-\frac{2 m r^2}{(q^2+r^2)^{3/2}}\right)^{-1}dr^2-r^2 d\Omega^2.
\label{eq:1.63}
\end{aligned} 
\end{equation}

\noindent The Kretschmann scalar is given by

 \begin{equation}
\begin{aligned}
K&=\frac{4a^2}{r^4}+\frac{16 a m}{r^2 \left(q^2+r^2\right)^{3/2}}
\\
&+\frac{12 m^2 \left(8 q^8-4 q^6 r^2+47 q^4 r^4-12 q^2 r^6+4 r^8\right)}{\left(q^2+r^2\right)^7}.
\label{eq:1.64}
\end{aligned}
\end{equation}

Calculating the limits when
${r\rightarrow 0}$ and 
${r\rightarrow \infty}$, we obtain the following results

 {\begin{equation}
\lim_{r\rightarrow 0}K=\infty\,\,\,\mbox{and}\,\,\,\lim_{r\rightarrow \infty}K=0.
\label{eq:1.65}
\end{equation}

\begin{figure}[h!]
  \centering
  \begin{subfigure}[b]{0.47\textwidth}
    \includegraphics[width=\textwidth]{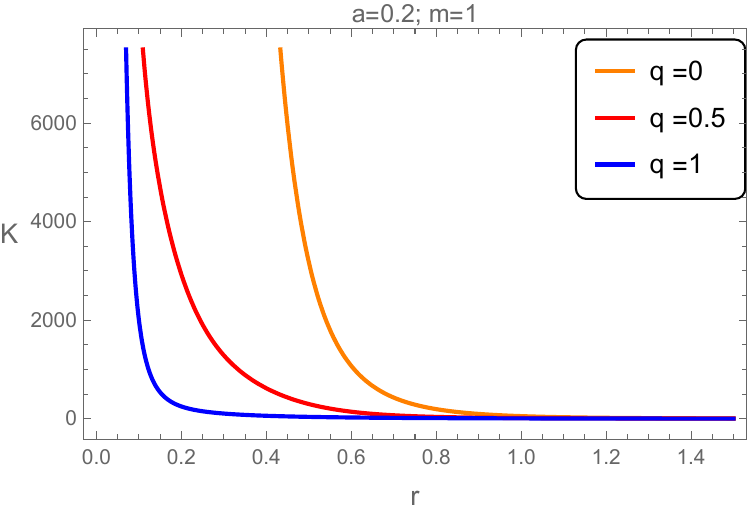}
    \caption{}\label{im6a}
  \end{subfigure}
  \vspace{.5cm}
  \begin{subfigure}[b]{0.47\textwidth}
    \includegraphics[width=\textwidth]{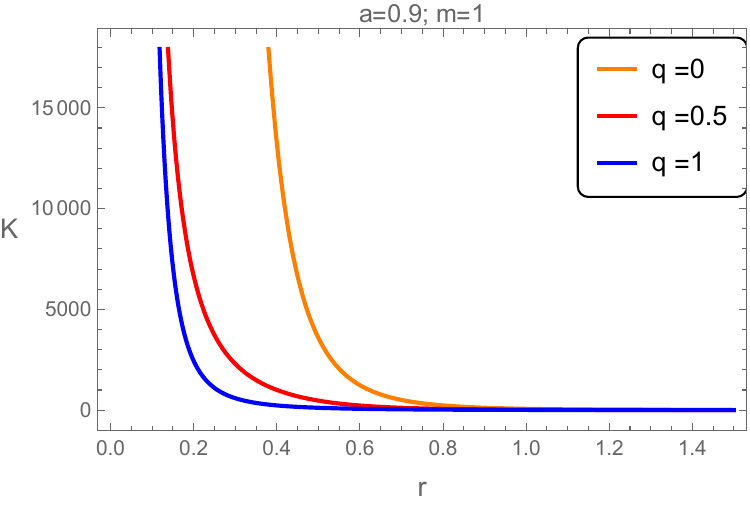}
    \caption{}\label{im6b}
  \end{subfigure}
\caption{Kretschmann scalar referring to the Bardeen black hole with a cloud of strings for different values of $q$ and $a$.}
  \label{im6}
\end{figure}

These results indicate that the presence of the cloud of strings surrounding the Bardeen black hole removes the regularity at the origin of the original Bardeen black hole.   The solution  is asymptotically flat, and this feature is illustrated in Fig. \ref{im6}.

\subsection{Bardeen black hole with cosmological constant and surrounded by quintessence}

The black hole's solution (Bardeen-Kiselev-AdS black hole) corresponding to this configuration can be obtained from 
Eq. (\ref{eq:1.31})
by assuming that  the parameters which codify the presence of a cloud of strings as well as of the electromagnetic field, vanish, namely
$a=0$ and $Q=0$}. 
Now, let us calculate the limits of the
Kretschmann scalar, when $r\rightarrow 0$
and $r\rightarrow \infty$.
For $\omega_q-2/3$, the solution is singular for $r\rightarrow 0$ and has a finite curvature for $r\rightarrow \infty$, according to what follows

\begin{equation}
\lim_{r\rightarrow 0}K=\infty\,\,\,\mbox{and}\,\,\,\lim_{r\rightarrow \infty}K=\frac{8 \Lambda ^2}{3}.
\label{eq:1.67.1}
\end{equation}

\noindent 
In the case where $\omega_q=-4/3$, the limit $r\rightarrow 0$ is given by 

\begin{equation}
\lim_{r\rightarrow 0}K=\frac{96m^2}{q^6}+\frac{32m\Lambda}{q^3}+\frac{8\Lambda^2}{3},
\label{eq:1.69.1}
\end{equation} which means that the solution is regular at the origin, for this value of $\omega_q$.

%

\section{Geodesics and effective potential}
\label{sec6}

We can model the trajectory of particles in regions close to a black hole by determining the geodesic
equations using the Lagrangian
which can be written as

\begin{equation}
\begin{aligned}
\mathcal{L}=\frac{1}{2}f(r)\dot{t}^2-\frac{1}{2}f(r)\dot{r}^2-\frac{r^2}{2}(\dot{\theta}^2 +\sin^2{\theta}\dot{\phi}^2),
\end{aligned}
\label{eq:1.76}
\end{equation}

\noindent where 
$f(r)$ is given by Eq. (\ref{eq:1.31}).
Note that we can define $L=2\mathcal{L}$, which for time-like geodesics, is equal to $+1$, for space-like geodesics, is equal to $-1$ and is equal to $0$ for null geodesics \cite{chandrasekhar1983mathematical}. 

Considering the Euler-Lagrange equations as follows

\begin{equation}
\frac{d}{d\tau}\left(\frac{\partial \mathcal{L}}{\partial\dot{x}^\mu}\right)-\frac{\partial \mathcal{L}}{\partial x^\mu}=0,
\label{eq:1.75}
\end{equation}

we can write the geodesic equations.

Taking $\mu=0$ and $\mu=3$ in Eq. (\ref{eq:1.75}), with $\mathcal{L}$ given by Eq. (\ref{eq:1.76}), we obtain, respectively:

\begin{equation}
\dot{t}=\frac{E}{\left(1-a-\frac{2 m r^2}{(q^2+r^2)^{3/2}}-\frac{\alpha}{r^{3 \omega _q+1}}+\frac{Q^2}{r^2}-\frac{1}{3}\Lambda  r^2\right)},
\label{eq:1.77}
\end{equation}

\begin{equation}
\dot{\phi}=-\frac{J}{r^2\sin^2\theta},
\label{eq:1.78}
\end{equation}

\noindent where $E$ and $J$ are constants of motion that can be interpreted as the energy and angular momentum of the particle moving near the black hole.

By setting $\theta=\pi/2$, we restrict the particle's motion to the equatorial plane of the black hole, so Eqs. (\ref{eq:1.77})-(\ref{eq:1.78}) reduce to: 

\begin{equation}
\dot{t}=\frac{E}{\left(1-a-\frac{2 m r^2}{(q^2+r^2)^{3/2}}-\frac{\alpha}{r^{3 \omega _q+1}}+\frac{Q^2}{r^2}-\frac{1}{3}\Lambda  r^2\right)},
\label{eq:1.79}
\end{equation}
\begin{equation}
\dot{\phi}=-\frac{J}{r^2},
\label{eq:1.80}
\end{equation}

\noindent  with $\dot{t}$ and $\dot{\phi}$ being the derivatives of $t$ and $\phi$ with respect to the proper time $\tau$. By inserting Eqs. (\ref{eq:1.79}) and (\ref{eq:1.80}) into Eq. (\ref{eq:1.76}), we arrive at

\begin{equation}
E^2=\dot{r}^2+V_{eff},
\label{eq:1.82}
\end{equation}
\noindent where
\begin{equation}
V_{eff}=f(r)\left(\frac{J^2}{r^2}+L\right).
\label{eq:1.83}
\end{equation}

\noindent This equation represents the effective potential for the geodesic motion in the charged Bardeen black hole with a cosmological constant and surrounded by quintessence and a cloud of strings, when $f(r)$ is given 
by Eq. (\ref{eq:1.31}), for the general case. 
Obviously, this general expression is also valid when we take into account an specific source or combination of them.

\noindent Using the relation

\begin{equation}
\begin{aligned}  
\left(\frac{dr}{dt}\right)^2\dot{t}^2=\dot{r}^2
\label{eq:1.85}
\end{aligned}
\end{equation}
\noindent in Eq. (\ref{eq:1.82}) and using Eq. (\ref{eq:1.83}) and Eq. (\ref{eq:1.79}), we obtain
\begin{equation}
\left(\frac{dr}{dt}\right)^2=f(r)^2\left[1-\frac{f(r)}{E^2}\left(\frac{J^2}{r^2}+L\right)\right].
\label{eq:1.86}
\end{equation}

\subsection{Radial movement of a massless particle}

For $J=0$ and $L=0$, we can write Eq. (\ref{eq:1.86}) as

\begin{equation}
\left(\frac{dr}{dt}\right)^2=f(r)^2.
\label{eq:1.87}
\end{equation}

By inserting Eq. (\ref{eq:1.85}) into Eq. (\ref{eq:1.87}), one obtains the relationship between the coordinates $t$ and $r$, which is given by

\begin{equation}
\pm t=\int\frac{1}{f(r)}dr.
\label{eq:1.88}
\end{equation}

From Eq. (\ref{eq:1.82}), one can derive the relation between the coordinate $r$ and the proper time $\tau$, namely,

\begin{equation*}
\left(\frac{dr}{d\tau}\right)^2=E^2,
\end{equation*}
\begin{equation}
\pm\tau=\frac{r}{E}.
\label{eq:1.89}
\end{equation}


\begin{figure}[h!]
  \centering
  \begin{subfigure}[b]{0.47\textwidth}
    \includegraphics[width=\textwidth]{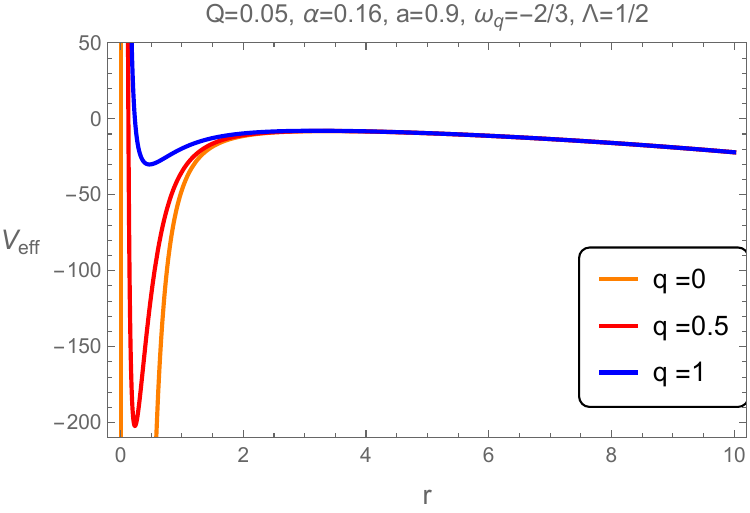}
    \caption{}\label{im7a}
  \end{subfigure}
  \vspace{.5cm}
  \begin{subfigure}[b]{0.47\textwidth}
    \includegraphics[width=\textwidth]{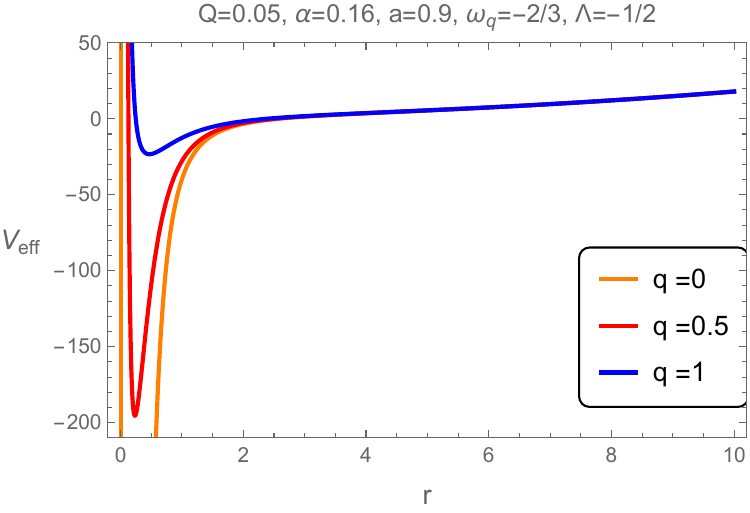}
    \caption{}\label{im7b}
  \end{subfigure}
   \begin{subfigure}[b]{0.47\textwidth}
    \includegraphics[width=\textwidth]{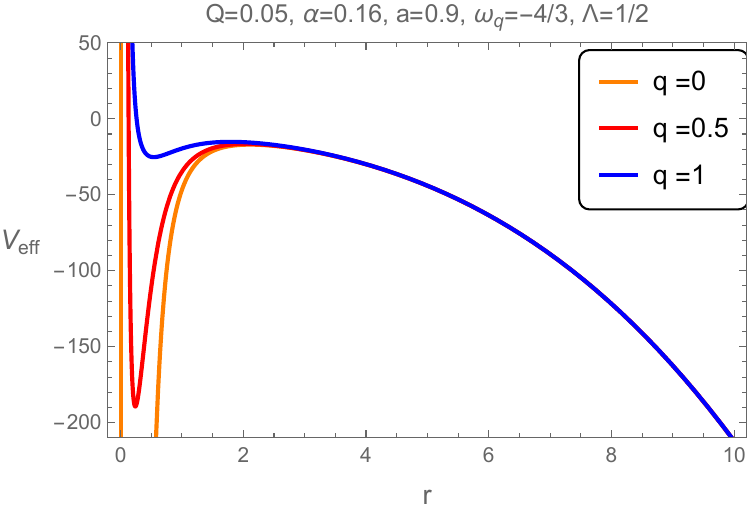}
    \caption{}\label{im7c}
  \end{subfigure}
   \begin{subfigure}[b]{0.47\textwidth}
    \includegraphics[width=\textwidth]{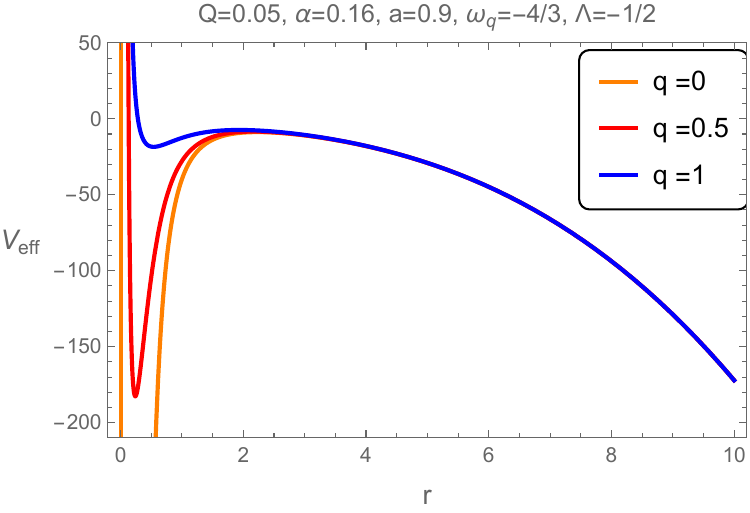}
    \caption{}\label{im7d}
  \end{subfigure}
 \caption{Effective potential for non-radial time-like geodesics ($L=1$ and $J^2=20$), for different values of $q$, $\omega_q$ and $\Lambda$.}
  \label{im7}
\end{figure}

\begin{figure}[h!]
  \centering
  \begin{subfigure}[b]{0.47\textwidth}
    \includegraphics[width=\textwidth]{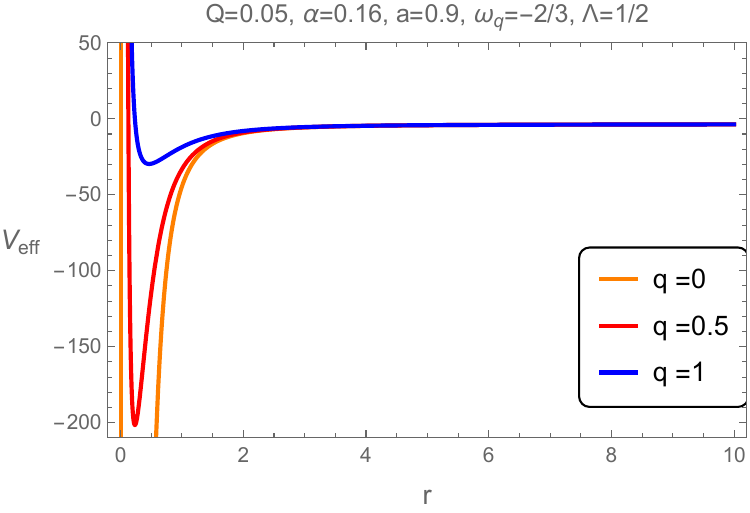}
    \caption{}\label{im8a}
  \end{subfigure}
  \vspace{.5cm}
  \begin{subfigure}[b]{0.47\textwidth}
    \includegraphics[width=\textwidth]{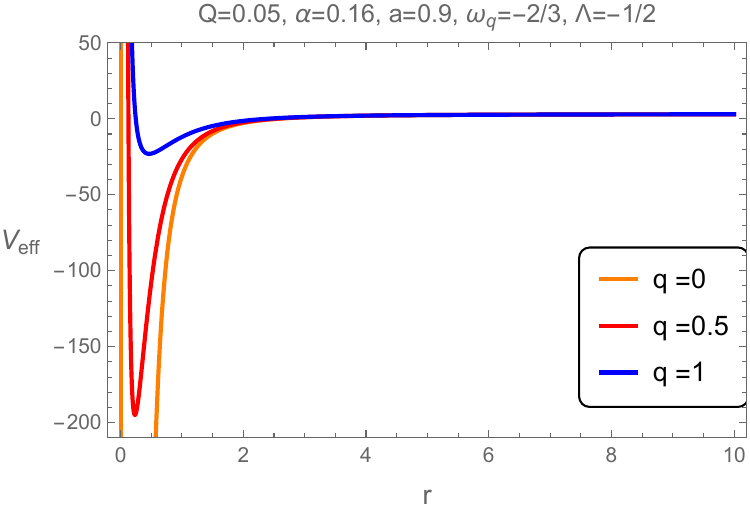}
    \caption{}\label{im8b}
  \end{subfigure}
   \begin{subfigure}[b]{0.47\textwidth}
    \includegraphics[width=\textwidth]{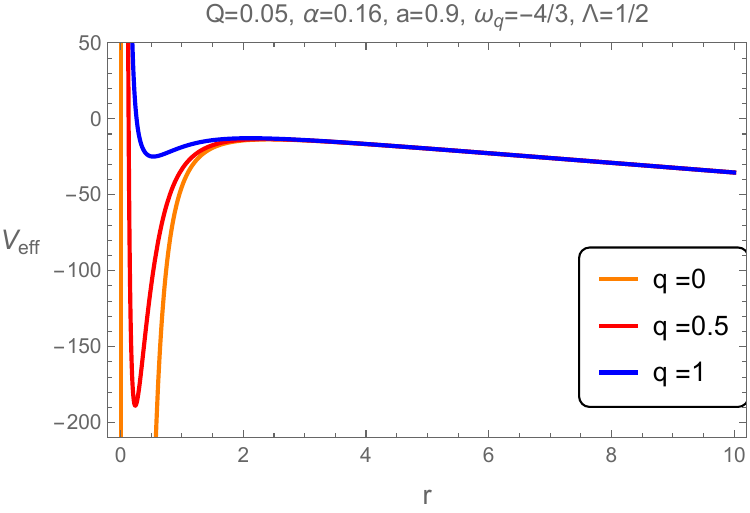}
    \caption{}\label{im8c}
  \end{subfigure}
   \begin{subfigure}[b]{0.47\textwidth}
    \includegraphics[width=\textwidth]{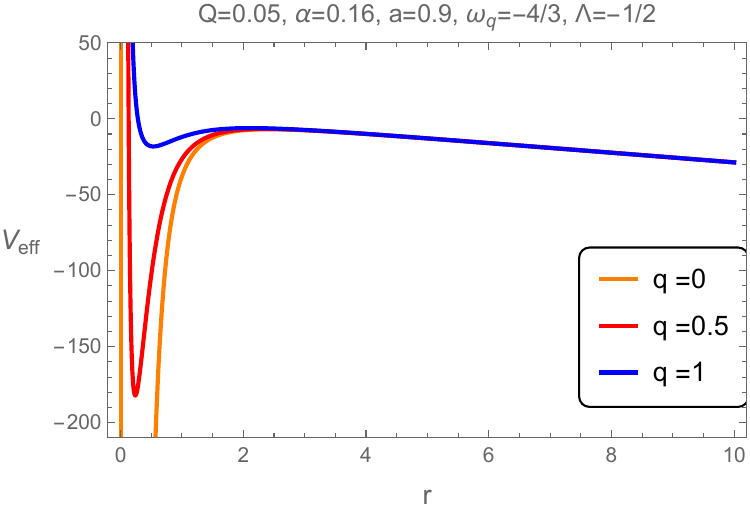}
    \caption{}\label{im8d}
  \end{subfigure}
 \caption{Effective potential for non-radial null-like geodesics ($L=0$ and $J^2=20$), for different values of $q$, $\omega_q$ and $\Lambda$.}
  \label{im8}
\end{figure}

\subsection{Radial movement of a massive particle}

Now let us consider the situation where $L=1$ and $J=0$. From Eq. (\ref{eq:1.86}), we get
\begin{equation}
\left(\frac{dr}{dt}\right)^2=f(r)^2-\frac{f(r)^3}{E^2}.
\label{eq:1.90}
\end{equation}

\noindent By inserting Eq. (\ref{eq:1.87}) into Eq. (\ref{eq:1.90}), the connection between $t$ and $r$ for the radial movement of the particle is constructed as

\begin{equation}
\pm t=\int\frac{dr}{\sqrt{f(r)^2-\frac{f(r)^3}{E^2}}}.
\label{eq:1.91}
\end{equation}

Using Eq. (\ref{eq:1.82}), we obtain the link between the proper time $\tau$ and the radial coordinate $r$:
\begin{equation*}
\left(\frac{dr}{d\tau}\right)^2=E^2-f(r),
\end{equation*}
\begin{equation}
\pm\tau=\int\frac{dr}{\sqrt{E^2-f(r)}}.
\label{eq:1.92}
\end{equation}

\subsection{Effective potential}

\begin{figure}[h!]
  \centering
  \begin{subfigure}[b]{0.47\textwidth}
    \includegraphics[width=\textwidth]{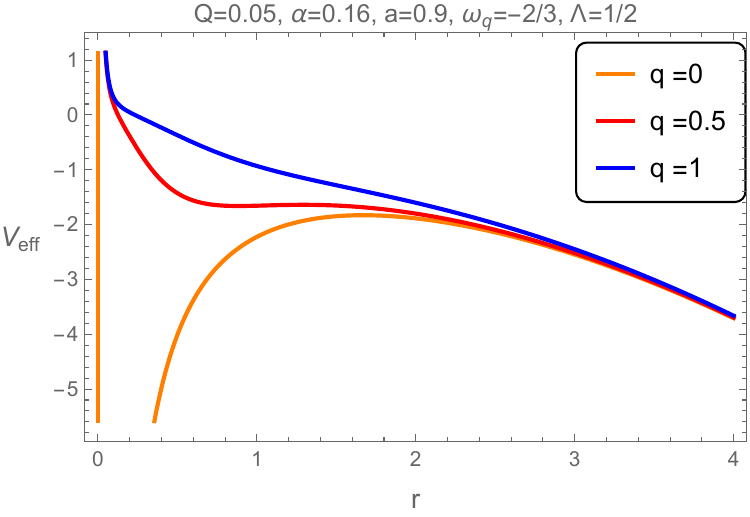}
    \caption{}\label{im9a}
  \end{subfigure}
  \vspace{.5cm}
  \begin{subfigure}[b]{0.47\textwidth}
    \includegraphics[width=\textwidth]{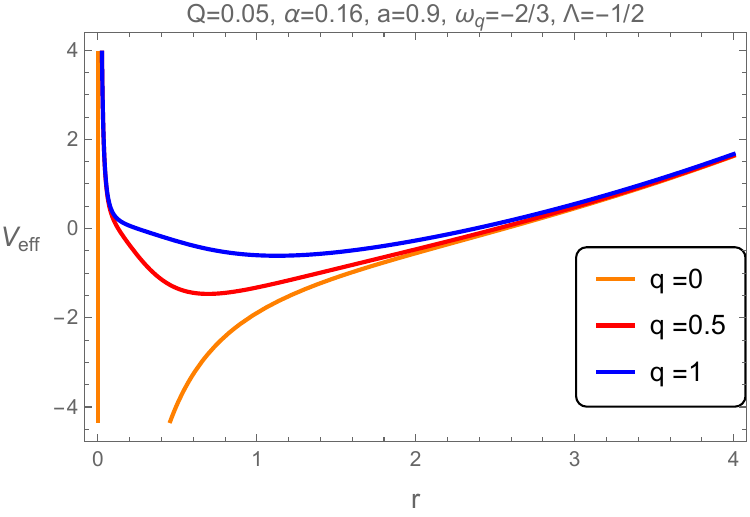}
    \caption{}\label{im9b}
  \end{subfigure}
   \begin{subfigure}[b]{0.47\textwidth}
    \includegraphics[width=\textwidth]{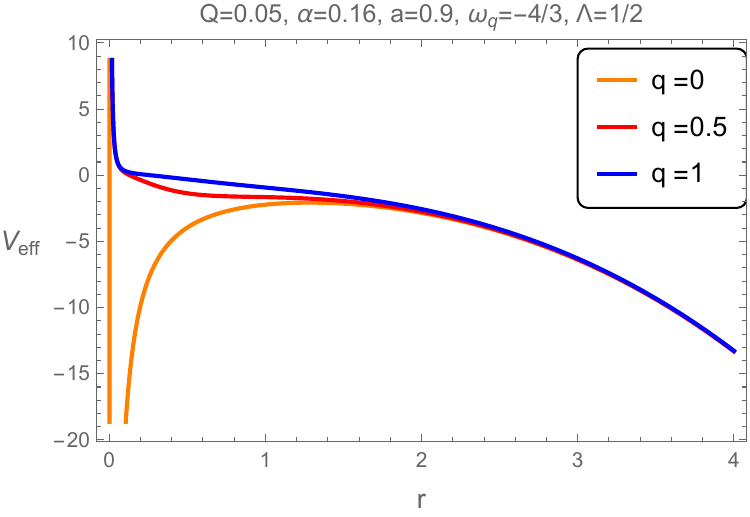}
    \caption{}\label{im9c}
  \end{subfigure}
   \begin{subfigure}[b]{0.47\textwidth}
    \includegraphics[width=\textwidth]{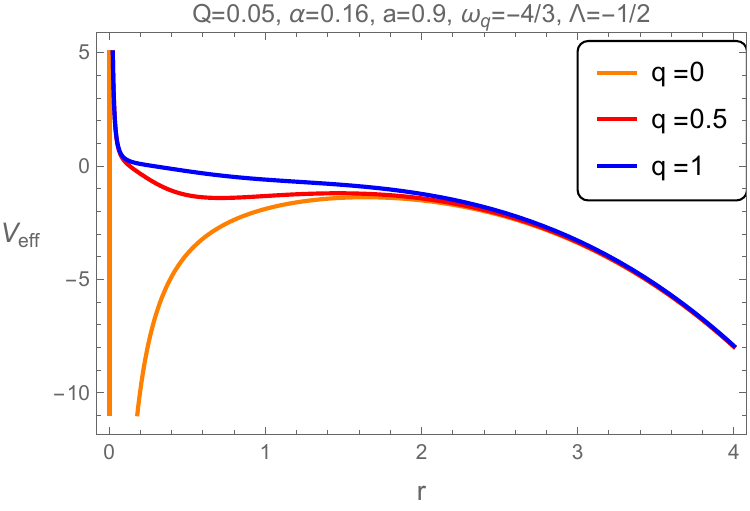}
    \caption{}\label{im9d}
  \end{subfigure}
 \caption{Effective potential for radial time-like geodesics ($J^2=0$ and $L=1$), for different values of $q$, $\omega_q$ and $\Lambda$.}
  \label{im9}
\end{figure}

\begin{figure}[h!]
  \centering
  \begin{subfigure}[b]{0.45\textwidth}
    \includegraphics[width=\textwidth]{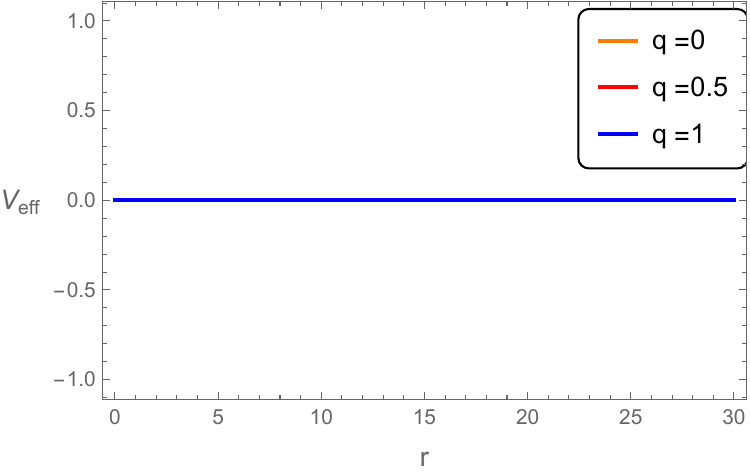}
      \end{subfigure}
  \caption{Effective potential for radial null-like geodesics  ($J^2=0$ and $L=0$),  for different values of $q$, $\omega_q$ and $\Lambda$.}
  \label{im10}
\end{figure}

The dynamics of geodesic motion is governed by the effective potential, 
$V_{eff}$, as given in Eq. (\ref{eq:1.83}). This potential provides insight into how massive and massless particles behave in the vicinity of the black hole. Consequently, Figs. \ref{im7}–\ref{im10} present $V_{eff}$ for various combinations of parameters $q$, $\omega_q$, and $\Lambda$,  illustrating both geodesics similar to time and null. Some panels focus on the region close to the horizon, where $V_{eff}$ is shown in more detail.

In Fig. \ref{im7}, we represent the effective potential for non-radial time-like geodesics ($L=1$ and $J^2=20$), for different values of $q$, $\omega_q$ and $\Lambda$. We can observe that, for $q=0$, there are no stable circular geodesics since the graphics do not show a local minimum for any value of $Q$, $\omega_q$, $\Lambda$, and $a$. On the other hand, for $q>0$, we can observe the possibility of the existence of stable circular geodesics, depending on the parameter values $Q$, $\omega_q$, $\Lambda$, and $a$. In the region near the black hole, $V_{eff}\rightarrow +\infty$. For regions far from the black hole, the effective potential also diverges. 

For non-radial null-like geodesics ($J^2=20$ and $L=0$), Fig. \ref{im8}, we can observe that, in all cases, $V_{eff}\rightarrow +\infty$ for regions near the black hole, $r\rightarrow 0$. For regions far from the black hole, $r\rightarrow \infty$, the $V_{eff}\rightarrow \pm 3.3333$ (Figs. \ref{im8a} e \ref{im8b}), and $V_{eff}\rightarrow -\infty$ (Figs. \ref{im8c} e \ref{im8d}). For $q=0$, there are no stable circular geodesics, since the graphics do not show local minima. The existence of stable circular orbits of massless particle around the black hole depends on $Q$, $\omega_q$, $\Lambda$, and $a$, as can be seen in Fig. \ref{im8}.

Concerning the behavior of radial time-like geodesics ($J^2=0$ and $L=1$), Fig. \ref{im9} shows us that the stability of radial movement does not occur for $q = 0$. On the other hand, for $q>0$, there will always be stable geodesics.

Ultimately, as shown in Fig. \ref{im10}, for radial null-like geodesics with $J^2=0$ and $L=0$, the effective potential remains constant and vanishes.

\newpage
\section{Concluding remarks}
\label{clonclusion}

The black hole solutions obtained which correspond to a generalization of the Bardeen black hole show the role played by each one of the parameters that codify the presence of the source. The analysis of regularity, from the point of view of Kretschmann's scalar calculus, shows that the obtained solutions do not have, in general, finite curvature at the origin.

In some specific cases, Kretschmann's scalar analysis shows regularity at the origin for Bardeen spacetime without extra sources, but only with a cosmological constant. For the Bardeen-Kiselev black hole, spacetime ceases to be regular for $\omega_q=-2/3$, but maintains its regularity for the phantom energy regime $\omega_q=-4/3$.
It is also observed that the inclusion of a cloud of strings or electric charge destroys the regularity of the metric, thus inserting a singularity at the origin. 

For non-radial time-like geodesics, there are no stable circular geodesics when $q=0$, while for $q>0$ the existence of these stable circular geodesics depends on parameters $Q$, $\omega_q$, $\Lambda$, and $a$. In the latter case, the effective potential diverges near and far from the black hole. 

For non-radial null-like geodesics, the effective potential diverges near the origin and is finite far from the black hole, for the following cases described in Figs. \ref{im8a} and \ref{im8b}.
For $q=0$, there are no stable circular geodesics, since the graphics do not show local minima. The existence of stable circular orbits of massless particle around the black hole depends on $Q$, $\omega_q$, $\Lambda$, and $a$, as can be seen in Fig. \ref{im8}.

Concerning the behavior of radial time-like geodesics, Fig. \ref{im9} shows us that for $q>0$, there will always be stable geodesics. Ultimately, as shown in Fig. \ref{im10}, for radial null-like geodesics, the effective potential remains constant and vanishes.

\backmatter
\bmhead{Acknowledgments}
V. B. Bezerra is partially supported by Conselho Nacional de Desenvolvimento Científico e Tecnológico - CNPq, Brazil, through the Research Project No. 307211/2020-7.

\section*{Declarations}
\textbf{Funding:} Conselho Nacional de Desenvolvimento Científico e Tecnológico - CNPq, Brazil. 

\noindent
\textbf{Conflict of interest/Competing interests:} The authors declare that they have no conflicts of interest.

\noindent
\textbf{Ethics approval and consent to participate:} Not applicable.

\noindent
\textbf{Consent for publication:} The authors consent with publication.

\noindent
\textbf{Data availability:} This study did not utilize underlying data.

\noindent
\textbf{Materials availability:} Not applicable.

\noindent
\textbf{Code availability:} Not applicable.

\noindent
\textbf{Author contribution:}
The authors contributed equally.

\bibliography{ref}
\end{document}